\documentclass{aa}  
\usepackage{txfonts}
\usepackage{graphicx}
\usepackage{epstopdf}
\usepackage{natbib} 
\bibpunct{(}{)}{;}{a}{}{,} 
\begin{document}
\title{Non-equilibrium calcium ionisation in the solar atmosphere}

\author{Sven Wedemeyer-B\"ohm\inst{1,2} \and Mats Carlsson\inst{1,2}}

\offprints{sven.wedemeyer-bohm@astro.uio.no}

\institute{Institute of Theoretical Astrophysics, University of Oslo,
  P.O. Box 1029 Blindern, N-0315 Oslo, Norway
  \and
  Center of Mathematics for Applications (CMA), University of Oslo,
  Box 1053 Blindern, N-0316 Oslo, Norway
}

\date{Received 24 November 2010; accepted XX January 2011}

\abstract{The chromosphere of the Sun is a temporally and spatially very varying 
medium for which the assumption of ionisation equilibrium is questionable.
}
{Our aim is to determine the dominant processes and timescales for the 
ionisation equilibrium of calcium under solar chromospheric conditions. 
}
{The study is based on numerical simulations with the RADYN code, which combines 
hydrodynamics with a detailed solution of the radiative transfer equation. 
The calculations include a detailed non-equilibrium treatment of hydrogen, calcium, 
and helium. 
Next to an hour long simulation sequence, additional simulations are produced, for 
which the stratification is slightly perturbed so that a ionisation relaxation 
timescale can be determined.  
The simulations are characterised by upwards propagating shock waves, which cause 
strong temperature fluctuations and  variations of the (non-equilibrium) ionisation 
degree of calcium. 
}{The passage of a hot shock front leads to a strong net ionisation of Ca~II,
rapidly followed by net recombination. 
The relaxation timescale of the calcium ionisation state is found to be of 
the order of a few seconds at the top of the photosphere and  10 to 30\,s in the 
upper chromosphere. 
At heights around 1\,Mm, we find typical values around 60\,s and in extreme cases 
up to $\sim 150$\,s. 
Generally, the timescales are significantly reduced in the wakes of 
ubiquitous hot shock fronts. 
The timescales can be reliably determined from a simple analysis of the eigenvalues 
of the transition rate matrix. 
The timescales are dominated by the radiative recombination from Ca\,III 
into the metastable Ca\,II energy levels of the 4d\,$^2$D term. 
These transitions depend strongly on the density of free electrons and therefore 
on the (non-equilibrium) ionisation degree of hydrogen, which is the main 
electron donor. 
}{
The ionisation/recombination timescales derived here are too long for the 
assumption of an instantaneous ionisation equilibrium to be valid and, on the other 
hand, are not long enough to warrant an assumption of a constant ionisation fraction.
Fortunately, the ionisation degree of \ion{Ca}{II} remains small in the height 
range, where the cores of the H, K, and the infrared triplet lines are formed. 
We conclude that the difference due to a detailed treatment of Ca~ionisation has 
only negligible impact on the modelling of spectral lines of \ion{Ca}{II} and the 
plasma properties under the conditions in the quiet solar chromosphere. 
}


\maketitle

\section{Introduction}
\label{sec:intro}

Modern observations unambiguously show that the chromosphere of the Sun is 
very dynamic and inhomogeneous  
\citep[see the reviews by, e.g.,][]{2001ASPC..223..131S, 2006ASPC..354..259J, 
2007ASPC..368...27R, 2009SSRv..144..317W}.
The interpretation of chromospheric observations, which often involves the 
construction of numerical models, thus must take into account spatial and temporal 
variations.  
Unfortunately, many simplifying equilibrium assumptions that can be made for the 
low photosphere, are not applicable for the layers above. 
A realistic chromosphere model must therefore account for deviations from the 
equilibrium state in the context of an atmosphere that changes on short timescales 
and small spatial scales. 
A successful example is the explanation of the formation of ``calcium grains'' by 
\citet{1997ApJ...481..500C}. 
Other examples of simulations with a time-dependent non-equilibrium treatment 
include the ionisation of hydrogen   
(\citeauthor{2002ApJ...572..626C} \citeyear{2002ApJ...572..626C}, 
hereafter referred to as Paper~I; 
\citeauthor{2006A&A...460..301L}
\citeyear{2006A&A...460..301L};
\citeauthor{2007A&A...473..625L}
\citeyear{2007A&A...473..625L})
and the concentration of carbon monoxide \citep{asensio03, 2005A&A...438.1043W}. 
When an individual process, be it a chemical reaction or atomic transition, changes 
on finite timescales longer than the dynamic timescales of the atmosphere, then it 
becomes necessary to solve a system of rate equations instead of the much simpler 
statistical equilibrium equations. 
In paper~I, it was shown that the resulting variation of the hydrogen ionisation 
degree is significantly reduced due to slow recombination rates. 
The ionisation degree then depends not only on the local gas
temperature, density, and the 
radiation field but also on the history of these properties.  
Already \citet{1991A&A...250..212R} considered the non-equilibrium ionisation 
of magnesium (\element{Mg}) and calcium (\element{Ca}) in 
their ab-initio solar chromosphere models. 
Their numerical implementation included a non-local thermodynamic equilibrium 
(NLTE) approximation with very simplified two-level atoms for 
\ion{Mg}{II}-{III} and \ion{Ca}{II}-{III}.

In this paper, we investigate the \ion{Ca}{II}-{III} ionisation balance in 
detail. 
The study is based on RADYN simulations similar to the hydrogen case in Paper~I.
The numerical simulations are described in Sect.~\ref{sec:sim}. 
The results of this study are presented in Sect.~\ref{sec:res}, 
followed by a discussion and conclusions in Sect.~\ref{sec:disc}.  

\section{Numerical simulations}
\label{sec:sim}

We use one-dimensional radiation hydrodynamic simulations calculated with 
RADYN, which are in most aspects very similar to the earlier simulation runs by 
\citet[][ paper~I]{1992ApJ...397L..59C, 1994chdy.conf...47C, 1995ApJ...440L..29C}. 
Here, we summarise only the most important properties. 
More details can be found in paper~I and references therein. 

The simulation code RADYN solves the equations of mass, momentum, energy, and 
charge conservation together with the non-LTE radiative transfer and population 
rate equations, on an adaptive mesh in one spatial dimension. 
It takes into account non-equilibrium ionisation, excitation, and radiative 
energy exchange from the atomic species H, He, and Ca with back-coupling on the 
hydrodynamics.  
Also, the effect of motion on the emitted radiation from these species is considered. 
Hydrogen and singly ionised calcium are modelled with six-level atoms and helium 
with a nine-level atom. 
Furthermore, doubly ionised helium is included. 
The transitions between all considered atomic levels are treated in detail. 
Each line is described with 31-101 frequency points, whereas 4-23 
frequency points are used for each continuum. 
Other elements than H, He, and Ca are taken into account in the form of background 
continua in LTE, which are derived with the Uppsala atmospheres program 
\citep{gustafsson73}.

\begin{figure}[t]
\begin{center}
\includegraphics[width=5cm]{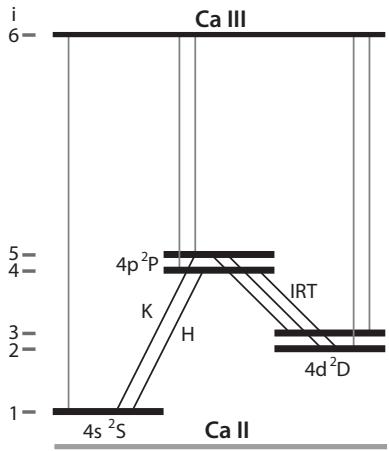}
\caption{Calcium model atom with the five lowest energy levels of \ion{Ca}{II}, 
a continuum level (\ion{Ca}{III}), and all included b-b and b-f transitions.
Energies are not to scale.} 
\label{fig:modelatom}
\end{center}
\end{figure}

The boundary conditions are the same as for the simulations in paper~I.
The lower and upper boundaries are both  transmitting. 
The lower boundary is located at a fixed geometrical depth, which corresponds 
to 480\,km below $\tau_{500} = 1$ in the initial atmosphere. 
We again use the same piston as in paper~I to excite waves at the bottom, which 
then propagate through the atmosphere. 
The piston velocity is based on Doppler-shift observations in a \ion{Fe}{I} line 
at $\lambda = 396.68$\,nm in the wing of the Ca\,H line \citep{1993ApJ...414..345L}. 
The upper boundary condition is located at a height of 10\,Mm. 
It represents a corona on top of the simulated layers 
with a temperature set to $10^6$\,K  and corresponding incident radiation 
\citep{1991JATP...53.1005T}.
The dynamic simulation has an overall duration of 3600\,s.

The numerical simulation is characterised by shock waves, which are excited as 
acoustic disturbances of small amplitude through the piston at the lower boundary. 
They propagate upwards and steepen into shocks with a saw-tooth profile above the 
photosphere. 
The simulated chromosphere is therefore subject to strong fluctuations in all 
hydrodynamic quantities. 
It takes a few 100\,s until the first shock waves transform the initially  
hydrostatic atmosphere into the characteristic dynamic atmosphere. 
The first 600\,s of the simulation are therefore excluded from the analysis.

\label{sec:runb}
A large number of additional short simulation runs are carried out, which are 
used for a timescale analysis (see Sect.~\ref{sec:timescales}). 
Each run starts from a snapshot of the detailed time-dependent (TD) simulation in 
the time window 600\,s to 2350\,s. 
The initial atmosphere of the individual runs contains the statistical 
equilibrium (SE) solution for the thermodynamic state in that snapshot. 
In the next time step, the atmosphere is perturbed by increasing the gas 
temperature by 1\,\% and the time-evolution is followed for 50\,s. 
In addition, we calculate the statistical equilibrium state of the perturbed 
atmosphere. 
For each snapshot in the dynamic simulation we thus have the following 
data: 
(i)~the statistical equilibrium solution of that snapshot, 
(ii)~the time-evolution after a perturbation has been applied, and 
(iii)~the final equilibrium state corresponding to the perturbed atmosphere.

\begin{figure}[t]
\begin{center}
\includegraphics{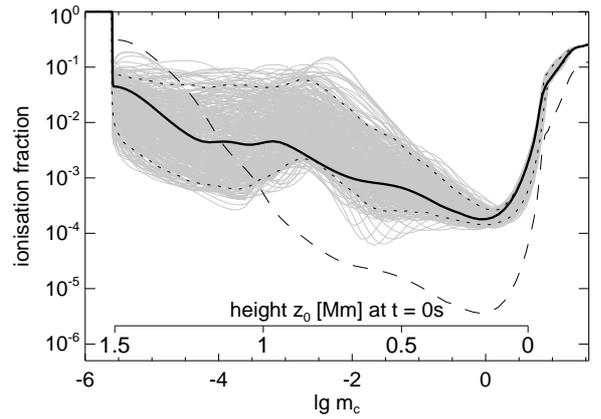}
\caption{Ionisation fractions of calcium ($\chi_{\element{Ca}}$, thick solid, see 
	Eq.~(\ref{eq:ionca})) and of hydrogen (dashed) as a 
	function of column mass in the initial radiative equilibrium atmosphere. 
	The height scale $z_0$ of the initial atmosphere is given as reference. 
	The \ion{Ca}{II}-{III} ionisation fraction rises from a minimum of $2\,10^{-4}$ 
	at \mbox{$\lg m_\mathrm{c} = 0.0$} ($z \approx 180$\,km) in the low photosphere 
	to $\sim 5$\,\% at the base of the transition region. 
	The grey lines represent $\chi_{\element{Ca}}$ in simulation snapshots 
	at an interval of 10\,s during the whole simulation sequence of 3600\,s. 
	The thin short-dashed lines are the 5\,\% and 95\,\% percentiles. 
  }
\label{fig:ionfrac}
\end{center}
\end{figure}
\begin{figure}[t]
\begin{center}
\includegraphics{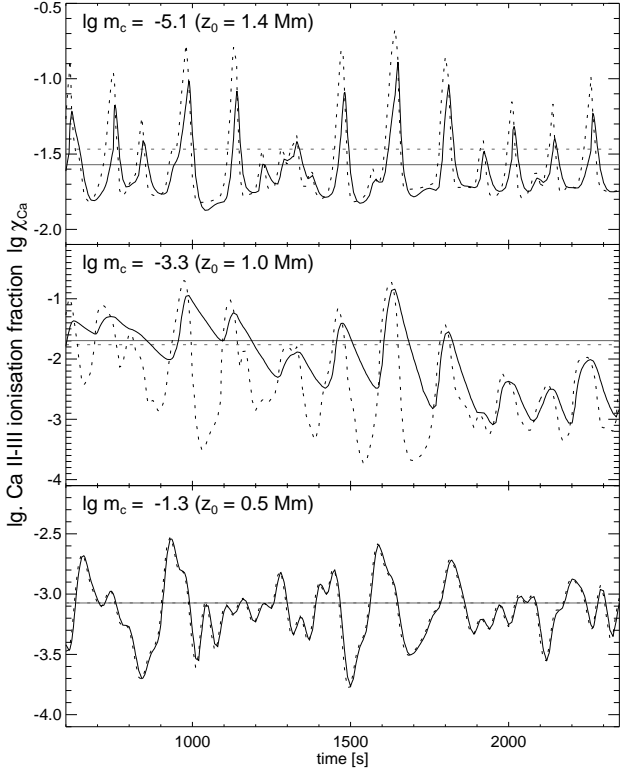}
\caption{Logarithmic \ion{Ca}{II}-{III} ionisation fraction as function of time at 
	different Lagrangian 
	locations. 
	The labels refer to the corresponding geometric height in the initial atmosphere.
	The lines represent the results from the non-equilibrium simulation (solid) and 
	for statistical equilibrium (dotted) calculated for the same values of the 
	hydrodynamic variables. 
	The horizontal lines are the corresponding time-averages. 
}
\label{fig:ionvstime}
\end{center}
\end{figure}

\subsection{The calcium model atom} 
\label{sec:modelatom}

The calcium model atom contains 5 bound states of singly ionised calcium 
(\ion{Ca}{II}) and a continuum level ($i = 6$), which represents the next 
ionisation stage (\ion{Ca}{III}).  
The lowest energy level with $i = 1$ is the ground state of 
\ion{Ca}{II} (4s\,$^2$S). 
We also include the two important energy level pairs of 
4p\,$^2$P ($i = 2, 3$)
and 
4d\,$^2$D ($i = 4, 5$). 
All five allowed radiative bound-bound (b-b) transitions are considered: 
the H and K resonance lines, and the infrared triplet (IRT). 
Also all five radiative bound-free (b-f) transitions with the corresponding 
photoionisation continua from the five lowest levels to level $i = 6$ 
are included. 
The model atom is illustrated in Fig.~\ref{fig:modelatom}. 

\section{Results}
\label{sec:res}

\subsection{Ionisation fraction}

In the solar chromosphere, calcium is mostly present in singly and doubly ionised 
form, i.e. as \ion{Ca}{II} and \ion{Ca}{III}, respectively. 
We therefore neglect neutral calcium and define 
the \ion{Ca}{II}-\ion{Ca}{III} ionisation fraction as the ratio of the population 
density of doubly ionised calcium $n_\mathrm{Ca\,III}$ and the population density 
of all included Ca ions 
($n_\mathrm{Ca,\,total} = n_\mathrm{Ca\,II} + n_\mathrm{Ca\,III}$). 
For our six level atom it reduces to
\begin{equation}
\label{eq:ionca}
\chi_{\element{Ca}} = 
\frac{n_\mathrm{Ca\,III}}{n_\mathrm{Ca\,II} + n_\mathrm{Ca\,III}}
= 
\frac{n_6}{\sum_{i=1}^6 n_i}\enspace,
\end{equation}
where $n_i$ is the population density of level~$i$ and $n_\mathrm{Ca\,III} = n_6$.
In the following, we refer to $\chi_{\element{Ca}}$ as \ion{Ca}{II}-{III} ionisation fraction. 
The variation of $\chi_{\element{Ca}}$ is shown in Fig.~\ref{fig:ionfrac} as a function of 
column mass density in the initial radiative equilibrium atmosphere (thick line). 
It has a minimum of $2\,10^{-4}$  in the middle photosphere  at 
\mbox{$\lg m_\mathrm{c} = 0.0$}, which corresponds to $z_0 \approx 180$\,km
on the height scale of the initial model. 
Like the hydrogen ionisation fraction (dashed line) it rises with height but 
less steeply. 
At the base of the transition region, $\chi_{\element{Ca}} \approx 5$\,\%  is reached.
The grey lines in Fig.~\ref{fig:ionfrac} illustrate the temporal evolution of 
the \ion{Ca}{II}-{III} ionisation fraction during the simulation. 
The thin dotted lines are the 5\,\% and 95\,\% percentiles, which enclose the typical 
data range of $\chi_{\element{Ca}}$. 
While the temporal variation is small below the middle 
photosphere, the fluctuations cover usually two orders of magnitude in the layers 
above. 
This is caused by the upward propagating shock waves.  

In Fig.~\ref{fig:ionvstime}, the \ion{Ca}{II}-{III} ionisation fraction $\chi_{\element{Ca}}$ is 
shown as function of time for three different Lagrangian locations, i.e. at fixed 
column mass densities. 
The corresponding geometrical heights $z$ vary in time. 
For orientation, the labels in each panel specify the geometrical height $z_0$
in the initial atmosphere.
Please note that the fluctuations of the ionisation degree appear much larger  
at a fixed geometrical height compared to a fixed column mass density because the 
atmospheric stratification and with it the height of the transition region 
is influcenced by the propagating shocks.  

The ionisation degree in the time-dependent non-equilibrium simulation 
(TD, solid line in Fig.~\ref{fig:ionvstime}) is compared to the statistical 
equilibrium values (SE, dotted line in Fig.~\ref{fig:ionvstime}). 
Both are calculated for the same atmospheric states. 
The differences between both cases are hardly discernible in the lower panel, which 
refers to the top of the photosphere. 
It shows that the ionisation fraction does not deviate considerably from the 
statistical equilibrium values in the lower atmosphere.  
The deviation is somewhat larger but still moderate in the upper chromosphere 
at heights around $z \sim 1400$\,km (upper panel in Fig.~\ref{fig:ionvstime}). 
At times of minimal ionisation, the time-dependent and statistical equilibrium 
results are very similar. 
At phases of higher ionisation in connection with passing shock fronts, 
the time-dependent simulation gives slightly smaller peak values, which lag 
behind the SE solution by 5 to 10\,s. 
The time-average of the SE case is $\sim 27$\,\% larger than the TD result
(see horizontal lines). 

In the middle chromosphere at heights around $z \sim 1000$\,km 
(see middle panel of Fig.~\ref{fig:ionvstime}), the SE  
ionisation fraction (dotted line) varies over two orders of magnitude on time 
scales of a few tens of seconds. 
In contrast, the detailed time-dependent simulation produces a \ion{Ca}{II}-{III} 
ionisation fraction that cannot follow the rapid changes (solid line).
After a strong increase and a shock-induced peak, which is similar in both cases, 
the ionisation fraction decays so slowly in the TD simulation that already the next 
shock front has arrived before the  equilibrium value can be attained. 
As we shall see in Sect~\ref{sec:rates}, this behaviour is caused by a relatively 
slow net recombination process. 
There is also a general time difference of the order of 5 to 10\,s between the 
occurrence of the ionisation peaks in the time-dependent 
simulation compared to the SE approach.  
At heights around $z \sim 1000$\,km, the TD ionisation fraction nevertheless varies 
strongly between values of $\sim 10^{-3}$ and $\sim 10^{-1}$. 
Although the temporal evolution of $\chi_{\element{Ca}}$ is very different for the TD and SE case 
at those heights, the time-averages are relatively similar. 
The SE average is only $\sim 14$\,\% smaller than the TD result 
(see horizontal lines).  

\subsection{Timescales}
\label{sec:timescales}

In the following, we derive the relaxation timescale of the \ion{Ca}{II}-{III} 
ionisation fraction (i)~numerically from the temporal evolution of an initially 
perturbed atmosphere (Sect.~\ref{sec:numtimescale}) and (ii)~from the eigenvalue 
analysis of the rate matrices (Sect.~\ref{sec:eigenvaluetimescale}). 

\begin{figure}[t]
\begin{center}
\includegraphics{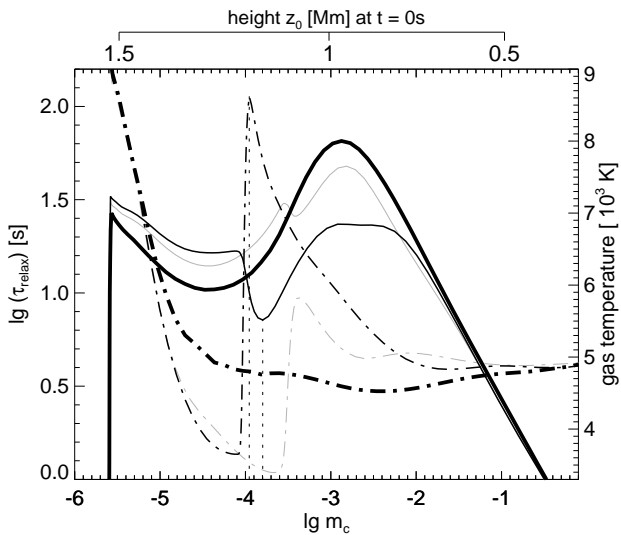}
\caption{Relaxation timescale (solid lines) and gas temperature (dot-dashed) 
	as function of column mass and time in the dynamic simulation. 
	The thick lines represent the relaxation timescale and the temperature 
	averaged in time over all time steps between $t = 600$\,s and $t = 2350$\,s. 
	Two individual time steps are drawn as thin lines: 
	$t = 1600$\,s (grey) and $t = 1620$\,s (black).  
	The dotted vertical lines mark the temperature peak and the minimum timescale 
	for this time step.
	For reference, the height scale ($z_0$) of the initial state is given on top. 
}
\label{fig:timescale_zt}
\end{center}
\end{figure}

\subsubsection{Numerical relaxation timescale}
\label{sec:numtimescale}

The simulation runs are used to determine a numerical timescale
as function of time by analyzing each time step in the time range from 600\,s to 
2350\,s individually (see Sect.~\ref{sec:runb}). 
For each time step, the timescale is derived from the exponential relaxation of the 
\ion{Ca}{II}-{III} ionisation fraction towards the equilibrium state of the 
perturbed atmosphere. 
The time-averaged relaxation timescale is then calculated by 
averaging over the timescales of the individual time steps. 
This average timescale is shown in Fig.~\ref{fig:timescale_zt} 
as function of column mass (thick solid line). 
It is very small in the photosphere ($< 1$\,s) and increases 
strongly with height until a maximum with values of the order of 1\,min is reached 
at heights around $z = 900$\,km.
The individual time steps show exactly the same behaviour in the photosphere, 
whereas the maximum in the chromosphere can reach values of up to 150\,s
in some cases. 
%
In the middle chromosphere above, the timescale is decreasing towards a minimum for all 
time steps. 
The average timescale drops to $\sim 10$\,s at $z_0 = 1200$\,km. 
The individual time steps also show a minimum but the exact position varies owing to the 
passage of shock waves. 
Values down to only a few seconds are found. 
The timescale starts to rise again at even larger heights but gets very small 
at the high-temperature transition region.

A number of successive individual time steps are represented by thin lines in  
Fig.~\ref{fig:timescale_zt}.  
They demonstrate the response of the relaxation time scale (solid lines) to the passage 
of a shock wave (gas temperature as dot-dashed lines). 
At first glance, it seems as if the relaxation timescale and gas temperature are neatly 
anti-correlated, i.e., the shortest time scales are found at the positions with the 
highest gas temperatures.
This would be expected from high temperatures resulting in large ionisation rates 
and thus small timescales.

A closer look, however, reveals that the minimum timescale is found at a height 
slightly below the temperature peak. 
This can be seen at $t = 1620$\,s 
(thin black lines in Fig.~\ref{fig:timescale_zt}). 
The height difference is marked by dotted vertical lines. 
At the beginning of the shock wave passage in the photosphere, when it is still a 
disturbance with small temperature amplitude, the timescale at the position of the 
disturbance is mostly set by the thermodynamic state of the background atmosphere 
and thus the aftermath of previous shock waves. 
It is only at heights of $z > 900$\,km, where the shock has steepened 
enough to have a significant effect on the timescale. 
There, a local timescale minimum develops at the height of the  
temperature peak.
While the shock front continues to propagate upwards in the atmosphere, the peak 
temperature grows, resulting in shorter and shorter timescales. 
The timescale minimum starts to appear lower than the peak temperature. 
This height difference grows during the passage through the chromosphere 
until values of the order of $\Delta\,z \sim 100$\,km are reached. 
From the time at which the locations of the temperature peak and the timescale minimum 
coincide, the relaxation timescale drops from a few 10\,s to just a few seconds. 
This behaviour is commonly found for all shock events.

When looking at the data as function of time at a fixed geometric height, 
the height offset translates into a small time difference between the occurrence 
of the temperature peak and the minimum timescale. 
This time difference decreases in most cases with height from a few 10\,s around 
$z \sim 1000$\,km to only a few seconds around $z \sim 1300$\,km. 
There are examples where the time difference does not show such a height 
dependence but rather varies around values of just a few seconds. 
The time and height offsets are caused by the influence of the hydrogen 
ionisation on the recombination rates of calcium via the electron density. 
See the analysis of the transition rates in Sect.~\ref{sec:rates}
for more details. 

\subsubsection{Timescales derived from eigenvalue analysis}
\label{sec:eigenvaluetimescale}

\begin{figure}[t]
\begin{center}
\includegraphics{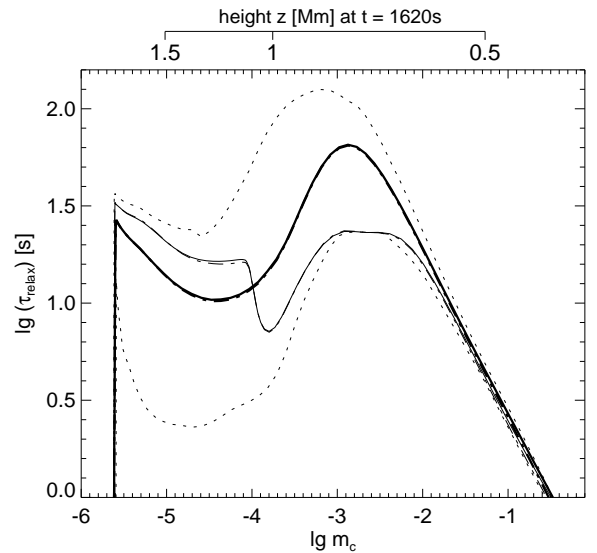} 
\caption{Numerical relaxation timescale as function of column mass determined 
	numerically (solid line) compared to the timescale determined from eigenvalues 
	of the rate matrix in the initial statistical equilibrium states (dot-dashed). 
	The thick lines represent the averages over all time steps between 
	$t = 600$\,s and $t = 2350$\,s, whereas the thin lines are for a 
	particular time step at $t = 1620$\,s. 
	Minimum and maximum values for both dynamic and eigenvalue timescales during 
	the considered time window are shown as grey dotted lines. 
	For reference, the height scale at $t = 1620$\,s is given on top.  
}
\label{fig:timescale_eigenv}
\end{center}
\end{figure}

As in paper~I, we use the eigenvalues of the transition rate matrix to 
determine the involved timescales and to 
indicate the processes governing the timescales. 
The rates in the dynamic simulation are affected by the previous 
history of the atmospheric state. 
The corresponding timescales are therefore not directly comparable to 
timescales derived from the numerical experiments in 
Sect.~\ref{sec:numtimescale}. 
Therefore, we use the statistical equilibrium state corresponding to
the hydrodynamic variables of a given snapshot
(see Sect.~\ref{sec:runb}).

The Lagrangian time-derivative of the population densities is
given by the rate equations:
\begin{equation}\label{eq:rate}
{D\vec{n}\over Dt}=\vec{P} \, \vec{n}
\end{equation}
with $\vec{n}$ being the vector of the level populations and $\vec{P}$ the 
rate matrix.
The matrix element $P_{ij}$ of the rate matrix is the transition rate per atom 
from level $i$ to level $j$ with \mbox{$P_{ii} = -\sum_j P_{ij}$}.  

The matrix elements depend on the population densities in general
through the non-linear radiation terms. If these non-linearities are
not dominant we may write the solution to 
Eq.~(\ref{eq:rate}) in
terms of eigenvectors and eigenvalues of the rate matrix
\citep[Paper~I;][]{2005JQSRT..92..479J}. 
For the Ca model atom (Sect.~\ref{sec:modelatom}) with 5 bound states and 
a continuum level (\mbox{$i = 6$}, \ion{Ca}{III}), the solution for the 
relaxation process can then be written as 
\begin{equation}
\vec{n} = \sum_{i=1}^{6}\, c_i \, \vec{v}_i \, e^{\lambda_i t}\quad.
\end{equation}
Here, 
$\vec{v}_i$ is the $i$th eigenvector of the rate matrix \vec{P}, 
$\lambda_i$ is the corresponding eigenvalue, and 
$c_i$ is a coefficient, which depends on the initial conditions. 
The equilibrium state corresponds to the eigenvector with a zero eigenvalue. 
All other eigenvalues are negative and represent the evolution of the level 
populations toward their equilibrium values. 
The longest timescale, on which the system relaxes towards the equilibrium state, 
is then given by the inverse of the smallest (in absolute value) nonzero eigenvalue.

In Fig.~\ref{fig:timescale_eigenv}, the timescales from the eigenvalue calculation 
are compared to the numerically determined relaxation timescales described in 
Sect.~\ref{sec:numtimescale}. 
The timescales averaged over all snapshots between $t = 600$\,s and 
$t = 2350$\,s match closely throughout the whole atmosphere. 
The eigenvalue and numerical timescales also match very well for the individual 
snapshots.
This shows that in the case of calcium ionization, the timescales do
not depend strongly on the non-linear parts of the rate matrix and we
have a convenient way of calculating the ionization/recombination
timescales from an eigenvalue analysis of the rate matrix.

\begin{figure*}[t]
\sidecaption
\includegraphics[width=12cm]{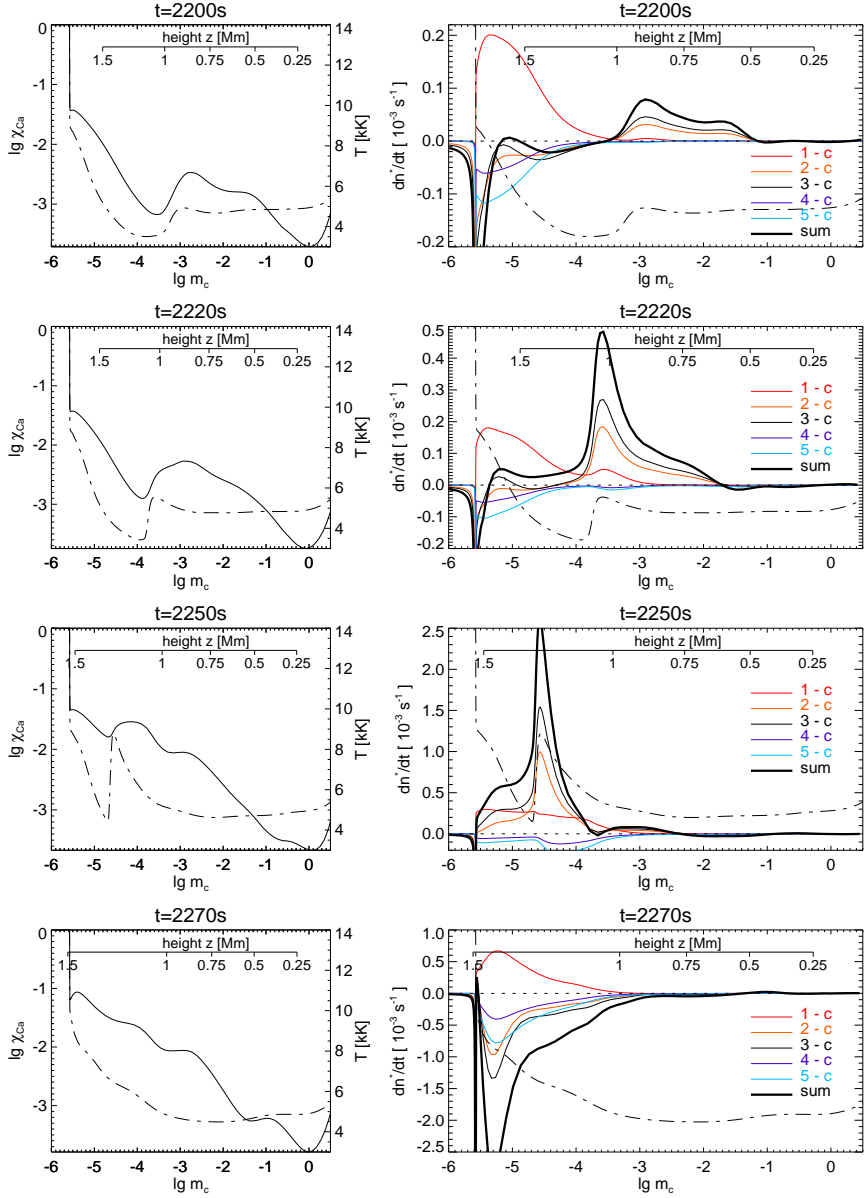}
\caption{\ion{Ca}{II}-{III} ionisation fraction (left column, solid line, left axis), gas 
temperature (left column, dot-dashed line, right axis), and normalised rates (right 
column) as a function of column mass for four time steps (rows from top to bottom) 
in the dynamic simulation. 
The rates for the radiative transitions from the bound levels~\mbox{1 - 5} to the 
continuum ($i = 6$, abbreviated ``c'') are divided by the total Ca number 
densities. 
Positive rates indicate ionisation, negative rates recombination. 
The sum of all radiative b-f transitions is represented by the thick solid line. 
The gas temperature is repeated in the right column as reference (dot-dashed line). 
The height axis for the individual time step is shown at the top of each panel. 
Please note the different data ranges for the individual time steps. 
}
\label{fig:ratesvsheight}
\end{figure*}

\subsubsection{Timescales of individual processes} 
\label{sec:timescale_process}

We now investigate which processes dominate the relaxation timescale by modifying the 
entries of the rate matrix prior to determining the eigenvalue timescale.   
The resulting timescales have qualitatively the same run with height but differ in 
their absolute value. 
Leaving out the collisional rates does not change the 
overall timescale in the middle chromosphere because the collisional transitions 
occur on timescales that are up to 5 orders of magnitude larger compared to the 
overall timescale. 
Also setting the radiative \mbox{b-b} transition rates to zero has no noticeable 
effect on the timescale. 
Setting the rates $R_{1\,6}$ and $R_{6\,1}$ for the radiative \mbox{b-f} 
transition between the ground level (\mbox{$i = 1$}) and 
the continuum (\mbox{$i = 6$}, Ca~III) to zero changes the timescale only slightly. 

We now set all matrix elements to zero except for the entries of individual \mbox{b-f} 
transitions. 
A rate matrix with non-zero entries only for $R_{1\,6}$ and $R_{6\,1}$ 
results in timescales of up to a few thousand seconds in the middle chromosphere. 
It implies that this transition, which connects the most populated energy level 
(\mbox{$i = 1$}) with the ionisation continuum, does not govern the overall timescale.  
It is the relaxation between level \mbox{$i = 3$} and the continuum, which produces 
the shortest timescales, followed by the transition between level \mbox{$i = 2$} 
and the continuum. 
Both transitions have timescales of the order of a few 10\,s. 
The ratio of the timescale with just a single transition and the timescale for the 
full rate matrix is about 2 for the transition 3-6 and $\sim 3$ for 2-6. 
The ratios for the transitions with lower level 4,5,1 are $\sim 20$, $~10$, and 
$\sim 70$, respectively.  
We conclude that the transitions between the metastable levels (2 and
3) and the continuum are the most important for setting the timescale
for ionization/recombination. Including only these transitions result
in a timescale that is only $\sim 18$\,\% longer than the timescale
derived from the full rate matrix.

\subsection{Transition rate analysis}
\label{sec:rates}

In contrast to the hydrogen case discussed in paper~I, 
the shortest timescales do not coincide exactly with the temperature peaks 
in the shock fronts (see Sect.~\ref{sec:timescales}). 
In this section, we show that this behaviour is caused by the dependence of the 
recombination rates on the density of free electrons, which is influenced by 
the ionisation/recombination timescale of hydrogen. 
For the Ca model atom considered here, the recombination process involves a 
collision between a free electron and a 
\ion{Ca}{III} ion \citep[see, e.g.,][ p.130f]{mihalas78}.
Calcium is only a minor electron donor in the solar atmosphere, whereas 
already small changes in the hydrogen ionisation fraction can lead to significant 
fluctuations of the electron density. 
As shown in paper~I, the hydrogen ionisation/recombination timescale 
is on average on the order of one to several hours in the chromosphere 
but can be strongly reduced in hot shock fronts. 
The timescales can then get as short as 10\,s to 20\,s. 
The resulting delayed release of electrons has a direct effect on the 
recombination rates of Ca so that the related timescales reach their 
minimum only shortly after the occurrence of the temperature peak.  

\begin{figure*}[t]
\sidecaption
\includegraphics[width=12cm]{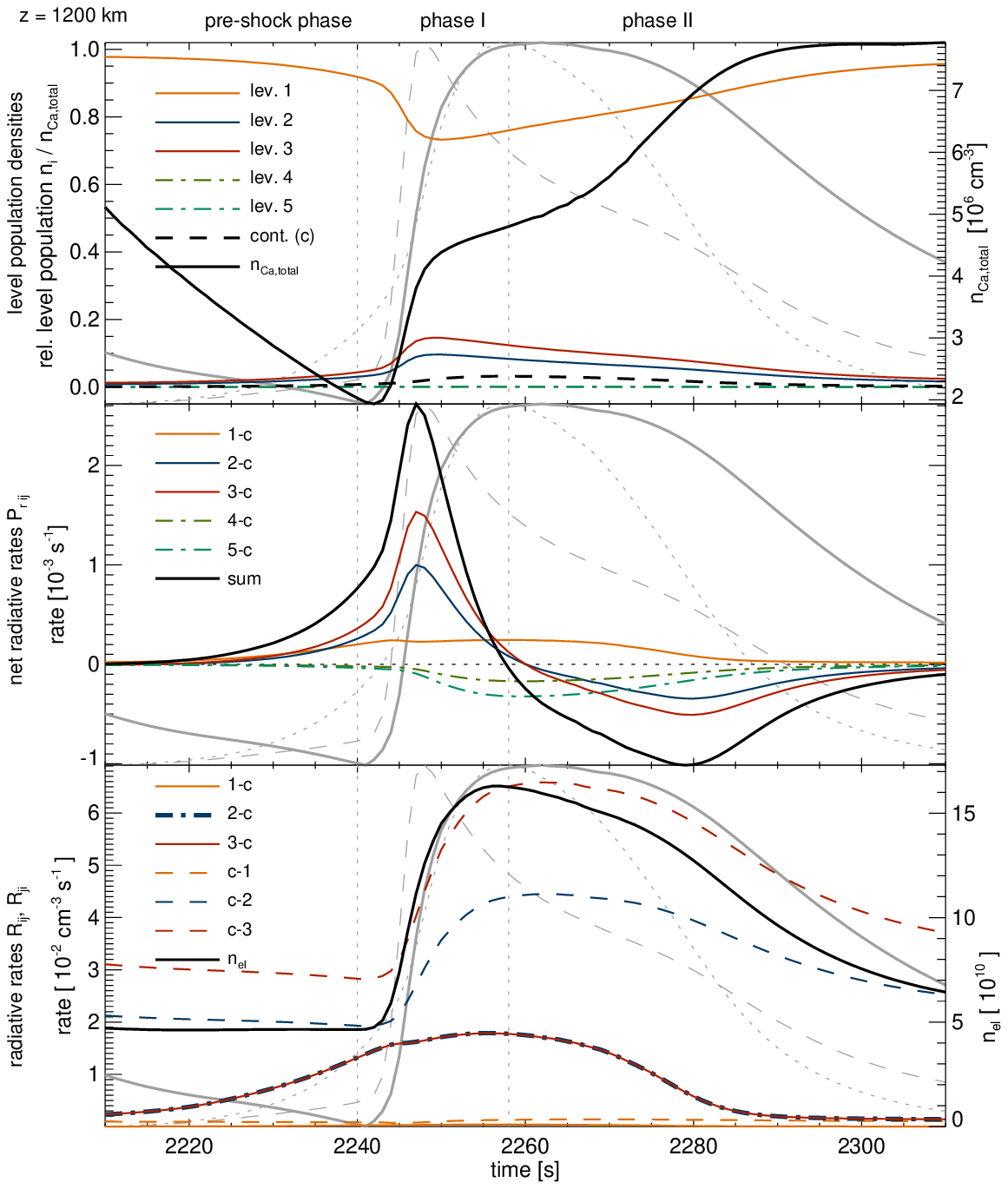}
\caption{Temporal evolution at a fixed geometric height of $z = 1200$\,km. 
\textit{Top:}~total Ca number density $n_\mathrm{Ca,total}$ and relative level 
populations $n_i / n_\mathrm{total}$. 
\textit{Middle:}~net radiative transition rates $P_{\mathrm{r}\,ij}$ for the 
b-f transitions between the bound levels and the continuum together 
with the overall sum $\sum_{i=0}^{5} P_{\mathrm{r}\,i6}$.
\textit{Bottom:}~the involved radiative rates $R_{ij}$ and $R_{ji}$ in comparison to 
the density of free electrons $n_\mathrm{el}$. 
The \ion{Ca}{II}-{III} ionisation fraction $\chi_{\element{Ca}}$ (grey dotted line), the gas 
temperature (grey dashed line), and the inversed relaxation timescale (grey solid 
line) are shown in all panels as reference. 
The rates for the transitions from the bound levels 1 - 5 to the continuum ($i = 6$)
are divided by the total Ca number densities. 
Positive rates indicate ionisation, negative rates recombination. 
}
\label{fig:rates_time}
\end{figure*}

In the following, the role of the individual ionisation and recombination processes 
is illustrated in detail for a typical shock that propagates through the model 
chromosphere at $t \sim 2200$\,s in the dynamic simulation. 
Most other shocks in the simulation show the same behaviour.  
In Fig.~\ref{fig:ratesvsheight}, the gas temperature, \ion{Ca}{II}-{III} ionisation 
fraction, and transition rates are shown as function of column mass density for the 
selected shock event. 
The four different time steps (rows from top to bottom) represent different stages of 
the upward propagating shock wave, which is most clearly seen as excess in gas 
temperature (dot-dashed line in the left column). 
The corresponding increase in ionisation fraction (solid line in the left column) is 
comparatively subtle with a peak (if discernible) slightly below the gas temperature 
peak.

The normalised net rate $P_{\mathrm{r}, i j}$ for a radiative transition between 
level~$i$ and level~$j$ is given by
\begin{equation}  
P_{\mathrm{r}, i j} = \frac{ n_i R_{ij} - n_j R_{ji} }{n_\mathrm{Ca, total}}
\label{eq:netrate}
\end{equation}
with the level population densities $n_i$ and $n_j$ and the radiative transition 
rates $R_{ij}$ (upwards, $i \rightarrow j$) and $R_{ji}$ (downwards, $j \rightarrow i$). 
The bound-free transition rates $P_{\mathrm{r}, i 6}$ are shown in the right 
column of Fig.~\ref{fig:ratesvsheight} together with the gas temperature as 
reference. 
The net rates $P_{\mathrm{r}, i j}$ are thus given per Ca~ion with positive values 
indicating net photoionisation and negative values indicating net recombination.  
Collisional b-f transitions have much smaller rates and can be neglected for this 
analysis. 
The sum of the radiative rates (thick solid) is close to zero in the lower parts 
of the model atmosphere ($\lg m_\mathrm{c} > -1.0$).
There, the rates are in or close to equilibrium with the rates of the individual 
transitions all being small. 
The sum of the rates is also close to zero in the upper parts of the model atmosphere 
below the transition region ($\lg m_\mathrm{c} \approx -5.6$) that have not yet been 
disturbed by the shock wave (e.g., $\lg m_\mathrm{c} < -3.5$ in the upper row).

In the shock front, the sum of the rates is increased, leading to net ionisation 
(positive values in the figure). 
The largest contribution is due to photoionisation from the meta-stable 
level~\mbox{$i = 3$}, followed by photoionisation from level~\mbox{$i = 2$}. 
As the lower levels of these transitions belong to the same term and are 
energetically close together (see Fig.~\ref{fig:modelatom}), the transition 
rates show a very similar behaviour. 
The photoionisation from the ground level ($i = 1$) is seen at all heights 
below the transition region with a local maximum in the shock front, where it 
has a small but positive contribution to the net ionisation. 
The normalised rate reaches much larger relative values above the shock front 
close to the transition region, compensating the recombination in most of the 
other b-f transitions.  
These large values are a result of the previous shock wave and long relaxation 
timescales for this transition (see Sect.~\ref{sec:timescale_process}).
The transitions from level \mbox{$i = 4$} and \mbox{$i = 5$} have much smaller 
rates with recombination prevailing throughout the model atmosphere. 
They both behave very similarly because the lower levels are energetically close 
together. 


As the shock propagates through the chromosphere, the normalised photoionisation 
rates increase significantly. 
While the sum of the rates (thick solid line) is of the order of $10^{-4}$\,s$^{-1}$ 
at the shock front at $z \approx 900$\,km (top row of Fig.~\ref{fig:ratesvsheight}), 
values of $\approx 2.5\,10^{-3}\,\mathrm{s}^{-1}$ are found at $z = 1150$\,km only 
50\,s later (bottom row). 
The second row from the bottom ($t = 2250$\,s) shows the instant shortly before 
the shock wave reaches the transition region (seen as sharp jump in gas temperature 
around $\lg m_\mathrm{c} = -5.6$).
Once the shock reaches the transition region (20\,s later, bottom row), the rates  
decrease rapidly. 
Except for a very small and smooth photoionisation from the ground level ($i = 1$), 
all other transitions and with it the sum of the rates show recombination in almost 
the entire model atmosphere from just behind the shock front down to the low 
photosphere.
It demonstrates that the timescales on which the ionisation equilibrium is restored 
are relatively short -- except for a comparatively sluggish photoionisation from 
the ground level.  


In Fig.~\ref{fig:rates_time}, the time evolution of the shock event is analysed 
at a fixed geometric height of \mbox{$z = 1200$\,km}. 
The vertical dotted lines divide the time span into a pre-shock phase,  
a phase~I (when the shock occurs) and a phase~II (during the passage of the shock 
wake). 
As the processes are complex and the timescales are small, the \ion{Ca}{II}-{III} 
ionisation fraction, the gas temperature, and the inverse relaxation timescale 
are shown in all panels for reference purposes (all scaled to their full data 
range). 
The gas temperature (grey dashed line) shows a clear shock signature with a 
rapid increase, a peak of $T_\mathrm{gas} = 8543$\,K at $t = 2248$\,s and a slower 
decay in the shock wake afterwards. 
The \ion{Ca}{II}-{III} ionisation fraction (grey dotted line) seems to react 
on the shock with some delay and also much smoother. 
It peaks only at $t = 2258$\,s with a value of 3.2\,\%, i.e. 10\,s after the 
occurrence of the temperature peak. 
The relaxation timescale $\tau$ reaches a minimum of 
$\tau_\mathrm{min} = 7.6$\,s even later at $t = 2262$\,s. 
The inverse timescale $\tau^{-1}$ is shown as solid grey line in the figure. 
While the change in phase~I occurs rapidly (with a slope of $\tau^{-1}$ 
being between those of the gas temperature and the ionisation fraction), 
the post-shock evolution in phase~II is much slower. 
Then (e.g. at $t = 2300$\,s) the timescale $\tau$ is still reduced while 
the gas temperature and the ionisation 
fraction are already relatively close to their pre-shock state again. 

The uppermost panel of Fig.~\ref{fig:rates_time} displays the total Ca~number density
$n_\mathrm{Ca,total}$, 
i.e. the sum of levels 1 to 6  (thick solid line, right axis). 
As the element abundance of Ca is constant in the simulation, it behaves exactly 
like the gas density. 
It decreases before the arrival of the shock front and rises rapidly again in  
the shock front and more gradually in the shock wake afterwards. 


The level population densities $n_i$, i.e. the number densities of Ca~atoms 
with their valence electron in a specific energy level~$i$, are divided by the 
total Ca number density $n_\mathrm{Ca, total}$ in the uppermost panel.  
It removes the fluctuations due to the change in gas density and reveals the 
relative distribution of the Ca atoms over the 6 included levels. 
At mid-chromospheric heights, usually 95\,\% or more of all \ion{Ca}{II} ions are 
in the ground level ($i = 1$) before and after the shock passage (i.e., in the 
cool phases). 
During the passage of a shock wave, the ground level is depopulated but usually 
maintains at least 70\,\% of $n_\mathrm{Ca,total}$ and 
thus remains the most populated level at all times. 
The population densities $n_i$ of all other levels with $i > 1$ increase 
during the shock passage as 
\ion{Ca}{II} atoms are excited from the ground level into a higher level. 
The increase of the population densities of levels 2 and 3 correlates with the 
change in gas temperature, while levels 4 and 5 react with a delay of 5\,s, shortly 
followed by the continuum level ($i = 6$).  
The density of the latter, $n_6$, increases at most to a few percent of $n_\mathrm{Ca,total}$. 
Most Ca atoms therefore remain singly ionised, even during the shock passages. 
The \ion{Ca}{III} ions are actually by a factor of more than 3 less 
abundant than the \ion{Ca}{II} ions with an electron in level 3 alone.  
Ionisation is thus only a ``minor process'' compared to the bound-bound transitions 
within the Ca atom.
The level populations of levels 4 and 5 on the other hand are much smaller than 
the  \ion{Ca}{III} number density. 
Even if all ions in these levels would ionise, the resulting rise of the 
\ion{Ca}{III} population would be negligible.
Levels 4 and 5 are nevertheless important as upper levels of radiative b-b 
transitions. 

\begin{figure}[t]
\begin{center}
\includegraphics{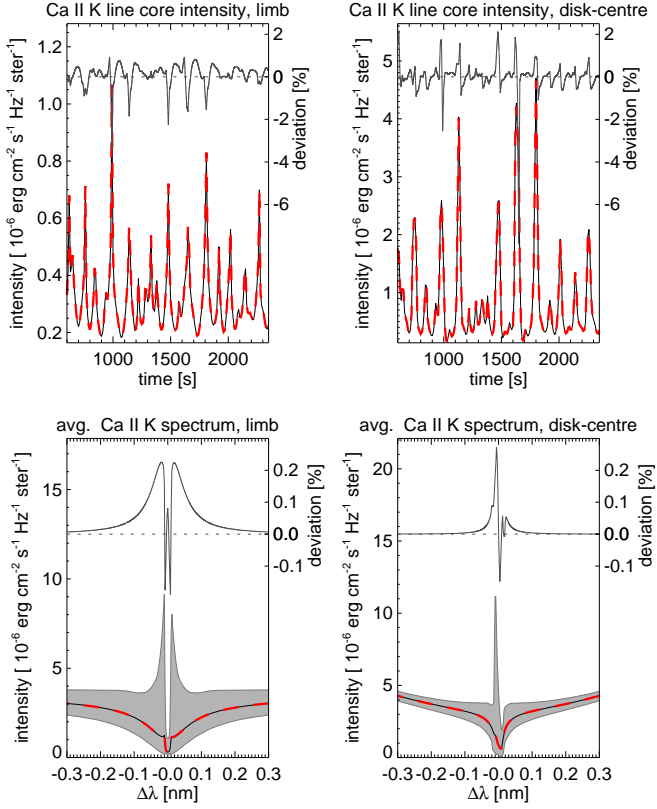}
\caption{
Emergent intensity in the \ion{Ca}{II}\,K line:
temporal evolution of the line core intensity (top row)
and time-averaged spectrum (bottom row) at the solar limb ($\mu = 0.05$, left column) 
and at disk-centre ($\mu = 1.0$, right column).  
Each panel contains the results from the non-equilibrium simulation 
($I_\mathrm{K, TD}$, solid) and from the statistical equilibrium 
solution ($I_\mathrm{K, SE}$, red thick dashed) together with the 
corresponding difference $(I_\mathrm{K, TD} - I_\mathrm{K, SE})/I_\mathrm{K, SE}$) 
between both (solid grey line at the top of each panel). 
The grey areas in the bottom panels represent the full data ranges of 
all intensity profiles during the simulation sequence between 
$t = 600$\,s and $t = 2350$\,s. 
}
\label{fig:intensity}
\end{center}
\end{figure}


The net rates $P_{\mathrm{r}, i j}$ for the radiative bound-free transitions are 
shown as function of time in the middle panel of Fig.~\ref{fig:rates_time}.  
They are again given per Ca~ion with positive values 
indicating net photoionisation and negative values indicating net recombination   
(see Eq.~(\ref{eq:netrate})). 
The sum of the net radiative rates $\sum_{i=1}^5 P_{\mathrm{r}, i 6}$ 
(thick solid line) gives the change of the \ion{Ca}{III} 
number density and thus the \ion{Ca}{II}-{III} ionisation fraction. 
Contributions from collisional b-f transitions are again small here. 
The sum exhibits a net \mbox{(photo-)}ionisation in phase~I, which peaks shortly 
(here 1\,s) before the gas temperature, and rapidly turns into net recombination 
for phase~II.  
While the shock wake is passing through the analysed height point, the 
recombination becomes weaker and the rates approach zero again. 
The b-f transitions from levels 3 and 2 contribute most (in that order) and 
show the same temporal behaviour as the sum. 
The net radiative bound-free rate for the ground level,  $P_{\mathrm{r},1 6}$,
exhibits net photoionisation throughout the whole shock passage. 
The rate increases with time and reaches a first hardly discernible maximum 
together with or shortly before the gas temperature. 
It roughly stays at the same level and smoothly declines in phase~II. 
At that time, it is the only transition with a positive net contribution. 
The bound-free transitions for levels 4 and 5 behave very much oppositely, 
producing a net recombination the whole time.  
The absolute values, however, stay small and even below the maximum recombination 
rates of $P_{\mathrm{r},2 6}$ and $P_{\mathrm{r},3 6}$ that are reached during 
phase~II. 


The peculiar behaviour of the net rates $P_{\mathrm{r},2 6}$ and 
$P_{\mathrm{r},3 6}$ is caused by a temporal delay between the photoionisation 
($n_j R_{ji}$) and the recombination part ($n_i R_{ij}$, cf. Eq.~(\ref{eq:netrate})). 
The photoionisation terms increase first and produce the strong net photoionisation 
in phase~I, while the recombination part reaches a maximum only about 20\,s later. 
At that time both parts are equal, resulting in a zero net rate. 
Afterwards the recombination part, which lags behind, stays larger than the
ionisation part, causing the net recombination in phase~II. 
The differences in the rates $R_{ij}$ are balanced by the number densities 
of the lower levels so that the recombination part is only slightly smaller than 
the photoionisation part. 
In contrast, the recombination into level $i = 1$ is much smaller than the 
corresponding photoionisation, which produces the smooth behaviour in time. 


Finally, the radiative bound-free transition rates $R_{i\,6}$ for  $i = 1, 2, 3$ 
are shown in Fig.~\ref{fig:rates_time}. 
The photoionisation rates $R_{2\,6}$ and $R_{3\,6}$, which are almost identical 
due to the similarity of their lower levels, exhibit no sharp response to the 
passage of the shock wave. 
In contrast, the recombination rates $R_{6\,3}$  and $R_{6\,2}$ increase strongly 
during the shock phase~I. 
The rate $R_{6\,3}$ is larger than $R_{6\,2}$ but both only differ by a 
constant factor. 
The strong increase of these rates correlates very well with the inverse 
relaxation time scale (grey solid line) and the electron density 
(solid black line). 
It demonstrates that the relaxation timescale is set by the recombination into the 
bound levels 3 and 2, which is directly depending on the density of free electrons. 
With hydrogen being the main electron donor in the chromosphere, 
the temporal behaviour of the recombination rates $R_{6\,3}$ and $R_{6\,2}$ is 
therefore similar to the \ion{H}{II} number density. 
This leads to the conclusion that the \ion{Ca}{II}-{III} ionisation in the quiet 
solar chromosphere is governed by the finite timescales on which hydrogen reacts 
on strong temperature changes. 

\section{Discussion and conclusions}
\label{sec:disc}

The ionisation/recombination timescales, which are here derived from 1-D RADYN 
simulations, are too long 
for the assumption of an instantaneous ionisation equilibrium to be
valid. On the other hand, the timescales are not long enough to
warrant an assumption of a constant ionisation fraction.
We find noticeable deviations from the ionisation equilibrium in the middle model 
chromosphere but  in general
the \ion{Ca}{II}-{III} ionisation fraction remains small.
The error due to the often made simplifying assumption of statistical equilibrium
is therefore  negligible for most applications. 
The effect is barely visible in the synthesized intensity for the 
diagnostically important spectral lines of \ion{Ca}{II}, i.e., 
the H and K lines and the infrared triplet. 
This finding is illustrated for the \ion{Ca}{II}\,K line 
in Fig.~\ref{fig:intensity}. 
The differences in the emergent intensity between the statistical 
equilibrium approach ($I_\mathrm{K, SE}$) and the time-dependent non-equilibrium simulation
($I_\mathrm{K, TD}$) are very small and hardly discernible in the temporal 
evolution of the intensity at a fixed wavelength position in the line 
core (top row) and even smaller (order of 0.1\,\%) in the
average spectrum (bottom row). 
The relative difference $(I_\mathrm{K, TD} - I_\mathrm{K, SE})/I_\mathrm{K, SE}$
reveals peaks with values of up to a few percent for wavelengths 
close to the line core. 
This is true both at disk-centre (right column) and close to the limb (left column). 
The differences are even smaller away from the line core and the emission peaks.
Noticeable deviations occur only in connection with shock fronts and thus sharp 
intensity increases. 
There, the effect is caused by a small temporal 
offset in the evolution of the level populations in the TD case with respect to the 
SE case, like 
it is seen for the ionisation fraction in Fig.~\ref{fig:ionvstime}.
It is safe to conclude that the time-dependent non-equilibrium treatment 
of the \ion{Ca}{II}-{III} ionisation appears to be of minor importance for the 
lower atmosphere in quiet Sun regions.

One restriction of the simulations is the use of only one spatial dimension, which 
could cause the shock wave profiles to be too extreme with potentially too high 
peak temperatures. 
In 3-D, the shock fronts are expected to be weaker owing to the larger number of 
degrees of freedom. 
The consequences of the deviations from the \ion{Ca}{II}-{III} ionisation 
equilibrium would be even less important in that case. 
Furthermore, we neglect the effect of incident radiation from the corona on 
the photoionisation of Ca. 
Most important in that respect would be the Lyman alpha line at 
$\lambda = 121.5$\,nm. 
However, the Lyman $\alpha$ photons have too little energy to ionise 
\ion{Ca}{II} from the ground state (11.88\,eV or $\lambda = 104.4$\,nm). 
The  photoionisation edges for levels 2 and 3, on the other hand, both lie 
close to the Lyman alpha line. 
The incident radiation certainly matters most at large heights close to the 
transition region. 
There, however, photoionisation from the ground level dominates, which is 
obviously not influenced by Lyman alpha photons.  
We conclude that the RADYN simulations employed here are sufficiently 
realistic for an evaluation of the importance of non-equilibrium 
effects for the ionisation of calcium in the context of a strongly varying 
chromosphere in quiet Sun regions. 
     
\begin{acknowledgements}
  This work was supported by a Marie Curie Intra-European Fellowship of the 
  European Commission (6th Framework Programme, FP6-2005-Mobility-5, Proposal 
  No.~042049) and a grant from the Research Council of Norway 
  (No.~191814/V30). 
\end{acknowledgements}
\bibliographystyle{aa} 

\begin{thebibliography}{19}
\expandafter\ifx\csname natexlab\endcsname\relax\def\natexlab#1{#1}\fi

\bibitem[{{Asensio Ramos} {et~al.}(2003){Asensio Ramos}, {Trujillo Bueno},
  {Carlsson}, \& {Cernicharo}}]{asensio03}
{Asensio Ramos}, A., {Trujillo Bueno}, J., {Carlsson}, M., \& {Cernicharo}, J.
  2003, \apjl, 588, L61

\bibitem[{{Carlsson} \& {Stein}(1992)}]{1992ApJ...397L..59C}
{Carlsson}, M. \& {Stein}, R.~F. 1992, \apjl, 397, L59

\bibitem[{{Carlsson} \& {Stein}(1994)}]{1994chdy.conf...47C}
{Carlsson}, M. \& {Stein}, R.~F. 1994, in Chromospheric Dynamics, ed.
  M.~{Carlsson}, 47

\bibitem[{{Carlsson} \& {Stein}(1995)}]{1995ApJ...440L..29C}
{Carlsson}, M. \& {Stein}, R.~F. 1995, \apjl, 440, L29

\bibitem[{{Carlsson} \& {Stein}(1997)}]{1997ApJ...481..500C}
{Carlsson}, M. \& {Stein}, R.~F. 1997, \apj, 481, 500

\bibitem[{{Carlsson} \& {Stein}(2002)}]{2002ApJ...572..626C}
{Carlsson}, M. \& {Stein}, R.~F. 2002, \apj, 572, 626

\bibitem[{{Gustafsson}(1973)}]{gustafsson73}
{Gustafsson}, B. 1973, Uppsala Astron. Obs. Ann., 5, 6

\bibitem[{{Judge}(2006)}]{2006ASPC..354..259J}
{Judge}, P. 2006, in Astronomical Society of the Pacific Conference Series,
  Vol. 354, Solar MHD Theory and Observations: A High Spatial Resolution
  Perspective, ed. J.~{Leibacher}, R.~F. {Stein}, \& H.~{Uitenbroek}, 259

\bibitem[{{Judge}(2005)}]{2005JQSRT..92..479J}
{Judge}, P.~G. 2005, \jqsrt, 92, 479

\bibitem[{{Leenaarts} {et~al.}(2007){Leenaarts}, {Carlsson}, {Hansteen}, \&
  {Rutten}}]{2007A&A...473..625L}
{Leenaarts}, J., {Carlsson}, M., {Hansteen}, V., \& {Rutten}, R.~J. 2007, \aap,
  473, 625

\bibitem[{{Leenaarts} \& {Wedemeyer-B{\"o}hm}(2006)}]{2006A&A...460..301L}
{Leenaarts}, J. \& {Wedemeyer-B{\"o}hm}, S. 2006, \aap, 460, 301

\bibitem[{{Lites} {et~al.}(1993){Lites}, {Rutten}, \&
  {Kalkofen}}]{1993ApJ...414..345L}
{Lites}, B.~W., {Rutten}, R.~J., \& {Kalkofen}, W. 1993, \apj, 414, 345

\bibitem[{{Mihalas}(1978)}]{mihalas78}
{Mihalas}, D. 1978, {Stellar atmospheres /2nd edition/} (San Francisco,
  W.~H.~Freeman and Co., 1978.~650 p.)

\bibitem[{{Rammacher} \& {Cuntz}(1991)}]{1991A&A...250..212R}
{Rammacher}, W. \& {Cuntz}, M. 1991, \aap, 250, 212

\bibitem[{{Rutten}(2007)}]{2007ASPC..368...27R}
{Rutten}, R.~J. 2007, in Astronomical Society of the Pacific Conference Series,
  Vol. 368, The Physics of Chromospheric Plasmas, ed. P.~{Heinzel},
  I.~{Dorotovi{\v c}}, \& R.~J. {Rutten}, 27

\bibitem[{{Schrijver}(2001)}]{2001ASPC..223..131S}
{Schrijver}, C.~J. 2001, in Astronomical Society of the Pacific Conference
  Series, Vol. 223, 11th Cambridge Workshop on Cool Stars, Stellar Systems and
  the Sun, ed. R.~J. {Garcia Lopez}, R.~{Rebolo}, \& M.~R. {Zapaterio Osorio},
  131

\bibitem[{{Tobiska}(1991)}]{1991JATP...53.1005T}
{Tobiska}, W.~K. 1991, Journal of Atmospheric and Terrestrial Physics, 53, 1005

\bibitem[{{Wedemeyer-B{\"o}hm} {et~al.}(2005){Wedemeyer-B{\"o}hm}, {Kamp},
  {Bruls}, \& {Freytag}}]{2005A&A...438.1043W}
{Wedemeyer-B{\"o}hm}, S., {Kamp}, I., {Bruls}, J., \& {Freytag}, B. 2005, \aap,
  438, 1043

\bibitem[{{Wedemeyer-B{\"o}hm} {et~al.}(2009){Wedemeyer-B{\"o}hm}, {Lagg}, \&
  {Nordlund}}]{2009SSRv..144..317W}
{Wedemeyer-B{\"o}hm}, S., {Lagg}, A., \& {Nordlund}, {\AA}. 2009, Space Science
  Reviews, 144, 317

\end{thebibliography}

\end{document}